\documentclass[9pt,twocolumn,twoside]{osajnl}
\graphicspath{{./fig/}}
\journal{ao} 
\usepackage{csquotes}
\usepackage{upgreek}
\setboolean{shortarticle}{false} 

\title{Ultra-compact reference ULE cavity}

\author[1,2]{Alexandre~Didier}
\author[1]{Jacques~Millo}
\author[1]{Baptiste Marechal}
\author[1]{Cyrus~Rocher}
\author[1]{Enrico~Rubiola}
\author[1]{Rom\'eo~Lecomte}
\author[1]{Morvan~Ouisse}
\author[3]{J\'er\^ome~Delporte}
\author[1]{Cl\'ement~Lacro\^ute}
\author[1,*]{Yann~Kersal\'e}

\affil[1]{FEMTO-ST Institute, univ. Bourgogne Franche-Comt\'e, CNRS, ENSMM, Besan\c{c}on, France}
\affil[2]{PTB, Bundesallee 100, 38116 Braunschweig, Germany}
\affil[3]{Centre National d'Etudes Spatiales (CNES), Toulouse, France.}

\affil[*]{Corresponding author: yann.kersale@femto-st.fr}

\dates{Compiled \today}

\ociscodes{(140.3425) Laser stabilization, (120.0120) Instrumentation, measurement, and metrology, (120.2230) Fabry-Perot}

\doi{\url{http://dx.doi.org/10.1364/XX.XX.XXXXXX}}

\begin{abstract}
We present a first experimental characterization of our ultra-compact, ultra-stable laser. The heart of the apparatus is an original Fabry-Perot cavity with a 25 mm length and a pyramidal geometry, equipped with highly-reflective crystalline coatings. The cavity, along with its vacuum chamber and optical setup, fits inside a 30 L volume. We have measured the cavity thermal and vibration sensitivities, and present a first estimation of the cavity fractional frequency instability at $\sigma_y(1\rm{s})=7.5{\times}10^{-15}$.
\end{abstract}

\setboolean{displaycopyright}{true}

\begin{document}

\maketitle

\section{Introduction}
High finesse Fabry-Perot cavities are widely used as frequency references to stabilize lasers sources. The improvement of these cavity-stabilized lasers are of prime importance for optical frequency standards \cite{ludlow2015}. Consequently, many laboratories are developing techniques to reduce the thermal noise of cavities, which now reaches the $10^{-17}$ decade, at the cost of very complex systems such as cryocoolers \cite{Kessler2012, Hafner2015} and active control of many technical parameters such as residual amplitude modulation, frequency fluctuations induced by the Doppler effect, or intensity fluctuations \cite{Zhang2014}.

In addition, these highly stable lasers are also used to synchronize optical frequency combs for ultra-low phase noise microwave signal generation \cite{Fortier2011, Didier2015, Xie2016}, with ultra-low frequency comb residual noise \cite{Xie2016, Bouchand2017}. Precision spectroscopy is also a field of interest. Such applications do not need state-of-the-art stabilized lasers, but compactness and simplicity are important criteria \cite{Davila-Rodriguez2017}. For this purpose, we are developing a laser based on a $25$~mm long cavity with a very compact vacuum chamber and an optical set-up reduced to the quintessential components.

In this article, we present our ultra-compact stabilized laser design, and present our first measurement of thermal and vibration sensitivities, as well as a preliminary estimation of the stabilized laser phase noise and fractional frequency stability.

\section{Description of the ultra-compact stabilized laser setup}
\subsection{Cavity design} \label{sec:1A}
\begin{figure}[h!]
	\centering
	\includegraphics[width=0.7\linewidth]{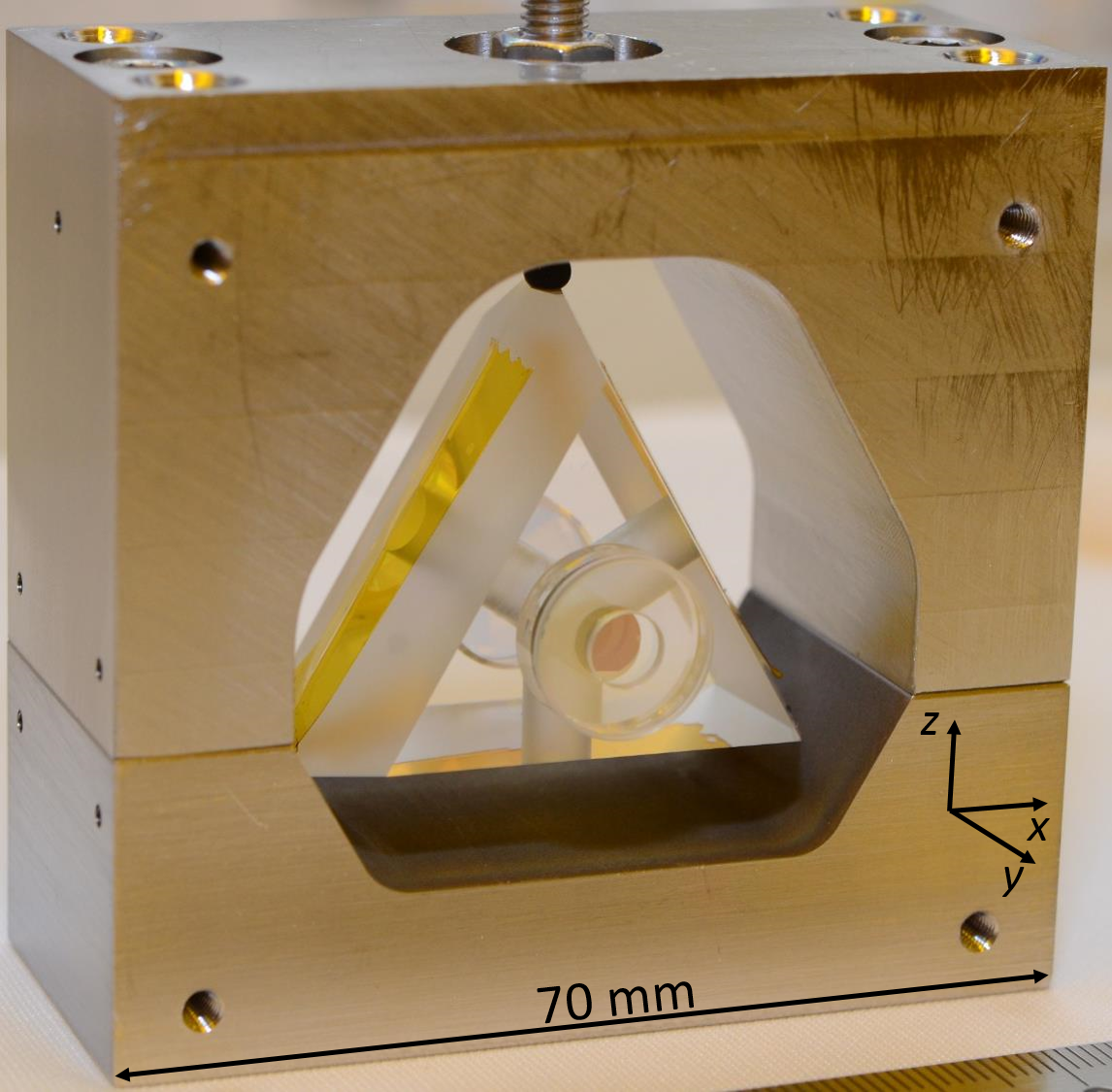}
	\caption{\textit{Double tetrahedral cavity in its stainless steel support. A Viton sphere on top of the cavity and two Viton pads at the bottom are used to fix the cavity in its support. }}
	\label{fig:cavity_pic}
\end{figure}

Several geometries have already been used to develop ultra stable Fabry-Perot cavities: horizontal and vertical cylinders, spherical and cubic spacers. Here we have developed a double tetrahedral geometry which is a good compromise between length, volume and thermal noise floor while keeping the overall cavity symmetry in order to reach very low residual vibration sensitivity. Our cavity is 25 mm long, corresponding to a free spectral range of 6 GHz, with a total spacer height of 36 mm. The cavity fits in an overall volume of 35~mm x 36~mm x 41~mm (Fig. \ref{fig:cavity_pic}). More details on the cavity design and the Finite Element simulations can be found in \cite{Didier2016}.

The tetrahedral spacer is machined in ultra-low expansion glass (ULE), while the cavity mirrors substrates are made of fused silicate. In order to reduce the coefficient of thermal expansion (CTE) zero-crossing shift due to the different CTE of ULE and fused silicate, ULE rings with dimensions of 12.7~mm outer diameter, 5~mm inner diameter and 1~mm thickness have been contacted to back of the cavity mirrors in order to obtain a CTE zero crossing near 11$^{\circ}$C \cite{Legero2010}.

The highly reflective coatings are made of crystalline GaAs/Al$_{0.92}$Ga$_{0.08}$As \cite{Cole2013}. These coatings exhibit lower mechanical losses than traditional dielectric coatings, thus reducing the cavity thermal noise floor. In our case, the use of crystalline coatings leads to a theoretical thermal noise floor around $1{\times}10^{-15}$, which is remarkable for such a short cavity. The cavity finesse at 1542~nm has been determined using the ringdown method. The 1/$e$ decay time of the exponential fit is 6.56 $\upmu$s, yielding an optical finesse of 247 000 for a mirror radius of curvature of 240~mm. This is to our knowledge the highest published finesse for an ultra-stable Fabry-Perot cavity based on crystalline coatings.


\subsection{Vacuum chamber}
\begin{figure}[h!]
	\centering
	\includegraphics[width=\linewidth]{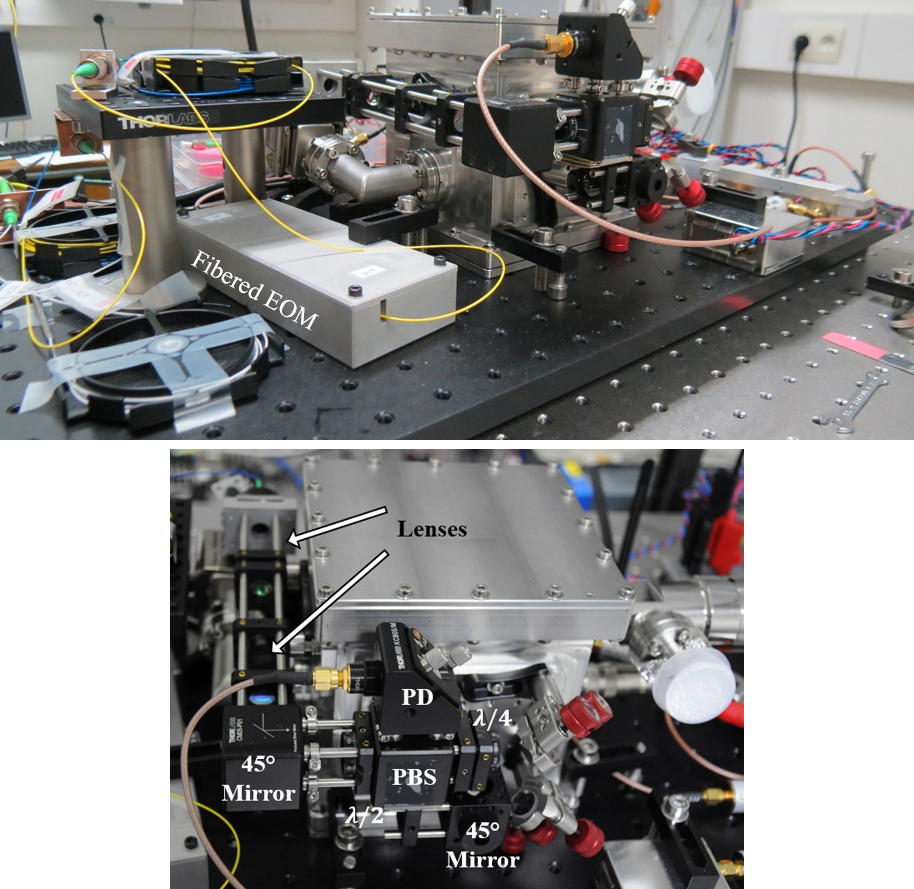}
	\caption{\textit{Optical setup. \emph{Top:} 30~cm $\times$ 60~cm breadboard, including the cavity in its vacuum chamber and all the optical setup except the laser source. \emph{Bottom:} close-up to the vaccum chamber with integrated optics for the light injection in the ultra compact ULE cavity. EOM: Electro-Optics Modulator. PD : PhotoDiode. PBS : Polarized Beam Splitter.}}
	\label{fig:optics}
\end{figure}

In order to reduce the setup footprint, a custom cubic vacuum chamber was designed and assembled. It allows for direct mounting of free-space optical elements to its walls, where breadboards have been included (see Fig. \ref{fig:optics}). Thermal fluctuations are mitigated using a passive stainless steel shield made with the cavity holder and surrounded by an additional active copper shield (see \cite{Didier2016} for details). Thanks to its high density, stainless steel indeed offers a high thermal capacitance for a given volume. The active temperature control is made using a Peltier element placed under the copper shield; the resulting temperature stability of the copper shield is below $2\times10^{-4}$ K between 1 s and $4\times 10^4$~s. The cavity response to a temperature step was fitted with a 1/$e$ response time of 20 000 s. The estimated attenuation of temperature fluctuations by the passive shield is around $100$~dB at 1~Hz.



\subsection{Optical setup}
\begin{figure}[h!]
	\centering
	\includegraphics[width=\linewidth]{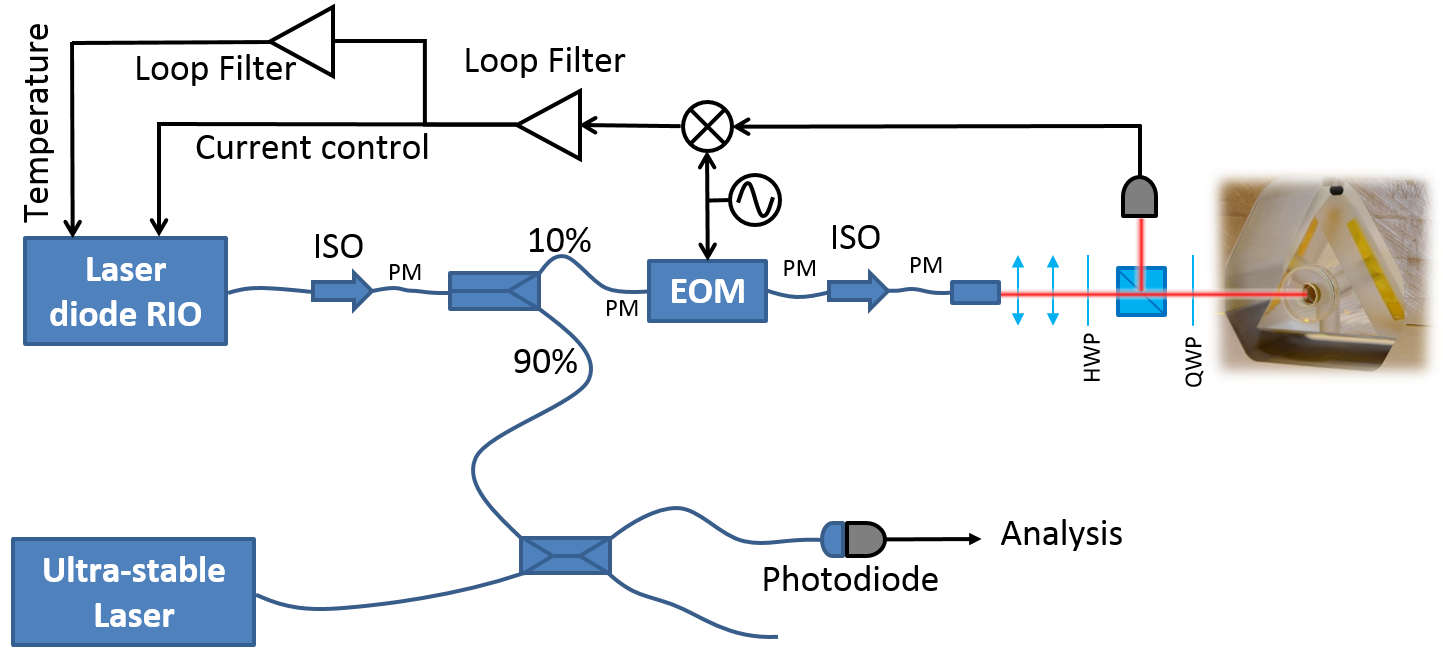}
	\caption{\textit{Schematic of the laser stabilized on the ultra compact cavity. All the components are fiber pigtailed except for the cavity mode matching (represented in red on the graph). HWP : Half Wave Plate. QWP : Quarter Wave Plate. PM: Polarization Maintaining. ISO: optical isolator.}}
	\label{fig:optics_schema}
\end{figure}

The optical setup has been boiled down to its quintessential elements. Only the mode-matching optics are free-space and based on 1/2" optical elements (see Fig. \ref{fig:optics_schema}). The rest of the optical set-up is based on pigtailed components with polarization maintaining fiber for compactness. The laser source is a compact, fibered extended cavity diode laser \footnote{RIO Planex} at 1542 nm. Its is packaged in a pigtailed butterfly case that we integrated in a homemade electronic control rack, including a low-noise current source with an integrated current noise of 658 nA between 10~Hz and 100~kHz and a digital temperature controller. A 90/10 optical coupler is used to extract the ultra-stable signal. A fibered electro-optical modulator (EOM) connected just before the output collimator provides phase modulation for the Pound-Drever-Hall stabilization method.

\section{Ultra-stable laser characterization}

\subsection{Vibration sensitivities}

Our simulations have predicted very low acceleration sensitivities for the double-tetrahedral geometry of our spacer. We have developed a dedicated setup for a well-controlled measurement of accelerations. The ultra-stable laser, including the vacuum chamber and the optical setup described in the previous section, is mounted on a thin aluminum breadboard supported by sorbothane posts. This apparatus is placed on an active vibration isolation platform, which can be modulated at frequencies between 0,1~Hz and 200~Hz in all directions. The platform is in turn placed on an optical table which provides passive isolation from ground vibrations.

The acceleration sensitivity coefficients $k_{x,y,z}$ relate the cavity fractional frequency noise $S_y$ to the acceleration noise $S_a$ so that \[S_y= \sum_{i=x,y,z} k_i^2 S_{a_i}\] where $y$ is along the cavity optical axis, $z$ is along the vertical direction, and $x$ is along the remaining transverse direction. We estimate $k_{x,y,z}$ by applying a sinusoidal modulation to the active isolation platform allong direction $i$ and measuring simultaneously the acceleration noise spectrum $S_{a_i}$ and the ultra-stable laser phase noise $S_{\phi}$. From measurements performed using accelerations of $10^{-4}\rm{m.s^{-2}}$ at 2 Hz and 6 Hz, we deduce the acceleration sensitivities listed in table \ref{table:acc}.

\begin{table}[b!]
\centering
{\begin{tabular}{|l|c|c|} \toprule
 Direction					& Measured at 2 Hz 					& Measured at 6 Hz 	\\ \midrule
 $x$								& $1,5 \times 10^{-11} / \rm{m.s^{-2}}$ &	$ 1,2 \times 10^{-11} / \rm{m.s^{-2}}$ 		\\
 $y$ 								& $8,7 \times 10^{-12} / \rm{m.s^{-2}}$	& $7,7 \times 10^{-12} / \rm{m.s^{-2}}$			\\
 $z$							 	& $1,1 \times 10^{-10} / \rm{m.s^{-2}}$	& $1,9 \times 10^{-10} / \rm{m.s^{-2}}$			\\ \bottomrule
\end{tabular}}
\caption{Acceleration sensitivities along the $x$, $y$ and $z$ measured at 2 Hz and 6 Hz.} 
\label{table:acc}
\end{table}

These sensitivities are much higher than our predictions. In particular, the sensitivity along the vertical axis is surprisingly high. In a quiet environment with a white acceleration noise of -95 dB(ms$^{-2})^2$/Hz, the vibration-noise contribution to the cavity fractional frequency instability would reach $1.4 {\times} 10^{-15}$, close to the thermal noise limit. In a noisier environment, the vibrations would become a dominant contribution to the cavity frequency noise.
Further investigation is needed to determine the origin of this high sensitivity. In particular, the absence of optical isolator in the free-space optical path implies that etalons might appear and cause an extra-sensitivity to acoustic noise. In a next study, we will try and eliminate these etalons by inserting an optical isolator in the free-space optical setup.

\subsection{Thermal expansion coefficient of the cavity}
As mentioned in section \ref{sec:1A}, the cavity was designed to exhibit a CTE turnover point at 11$^{\circ}$C. However, we found no inversion temperature when scanning the cavity setpoint between 8 $^{\circ}$C and 16 $^{\circ}$C. The cavity shows a linear sensitivity instead, with a relative thermal sensitivity of about $5{\times}10^{-8}$/K at 10 $^{\circ}$C. This sensitivity might stem from the influence of the stainless steel holder and the underestimated stiffness of the Viton balls. When designing our cavity, the influence of the holder was indeed not taken into account when calculating the CTE zero-crossing.

This thermal sensitivity yields short-term fluctuations not far below the thermal noise limit of the cavity, which we estimated at about $1{\times}10^{-15}$. Indeed, having measured the cavity response to a temperature step as well as the temperature temporal fluctuations at the copper shield, the resulting influence of thermal fluctuations to the fractional frequency instability is of $8.8{\times}10^{-16}\tau$ from 1 to 1000 s. A future upgrade of the setup will include an INVAR holder, which should allow us to recover a zero-CTE operating point and reach the cavity thermal noise.

\subsection{Laser phase noise and fractional frequency instability}
The stabilized laser phase noise and fractional frequency stability are measured by comparison with a telecom laser stabilized to a spherical cavity. This laser has been characterized independently by comparison with a cryogenic sapphire oscillator in the microwave domain, using optical frequency division techniques \cite{Didier2015}. Its frequency stability is better than that of the compact cavity under test.

\begin{figure}[h!]
	\centering
	\includegraphics[width=1\linewidth]{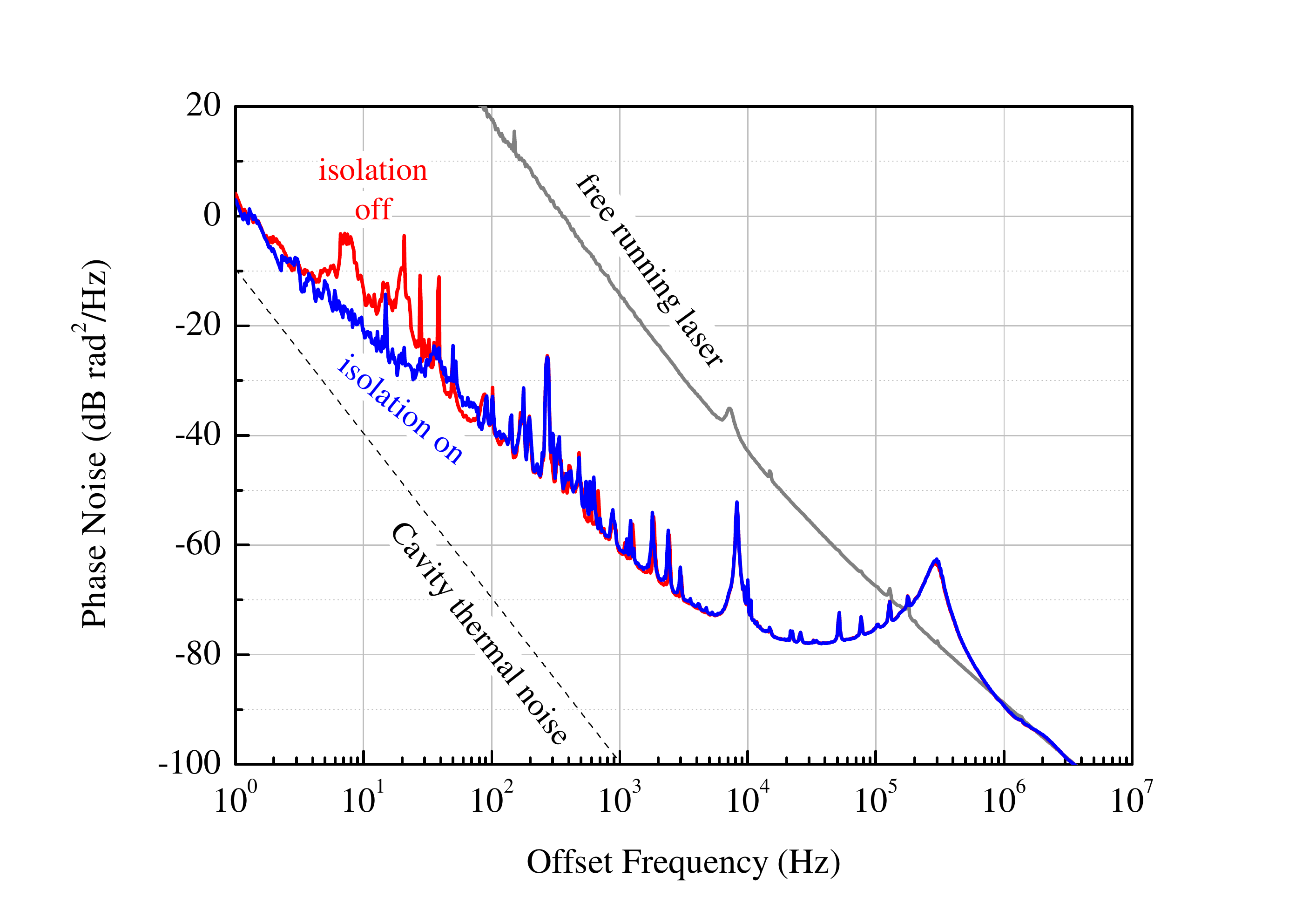}
	\caption{\textit{Phase noise of the laser locked to the compact cavity with (blue) and without (red) active vibration isolation, at 194 THz. We display the free-running laser noise (grey curve) and the cavity thermal noise limit (dashed line) for reference.}}
	\label{fig:phasenoise}
\end{figure}

We measure a phase noise of $3$~dBrad$^2 /$Hz at 1~Hz with a slope in $f^{-2}$ (see Fig. \ref{fig:phasenoise}). We can see the influence of the active vibration isolation platform between $\sim 4$~Hz and $\sim 60$~Hz;  at $1$~Hz, the phase noise is not impacted by the residual vibrations of the optical table. Assuming perfect frequency division to 10~GHz, the resulting phase noise would amount to -84~$\rm{dBrad}^{2}.\rm{Hz}^{-1}$.

The phase noise measured at $1$~Hz is $13$~dB higher than the thermal noise of the cavity. Moreover, the $f^{-2}$ slope between 3 Hz and 3 kHz denotes white frequency noise. We think that this noise might stem from our electronic frequency control loop, and are currently investigating the problem; lower electronic noise combined to higher servo gain should allow us to reach the cavity thermal noise in our current operating conditions.

\begin{figure}[h!]
	\centering
	\includegraphics[width=1\linewidth]{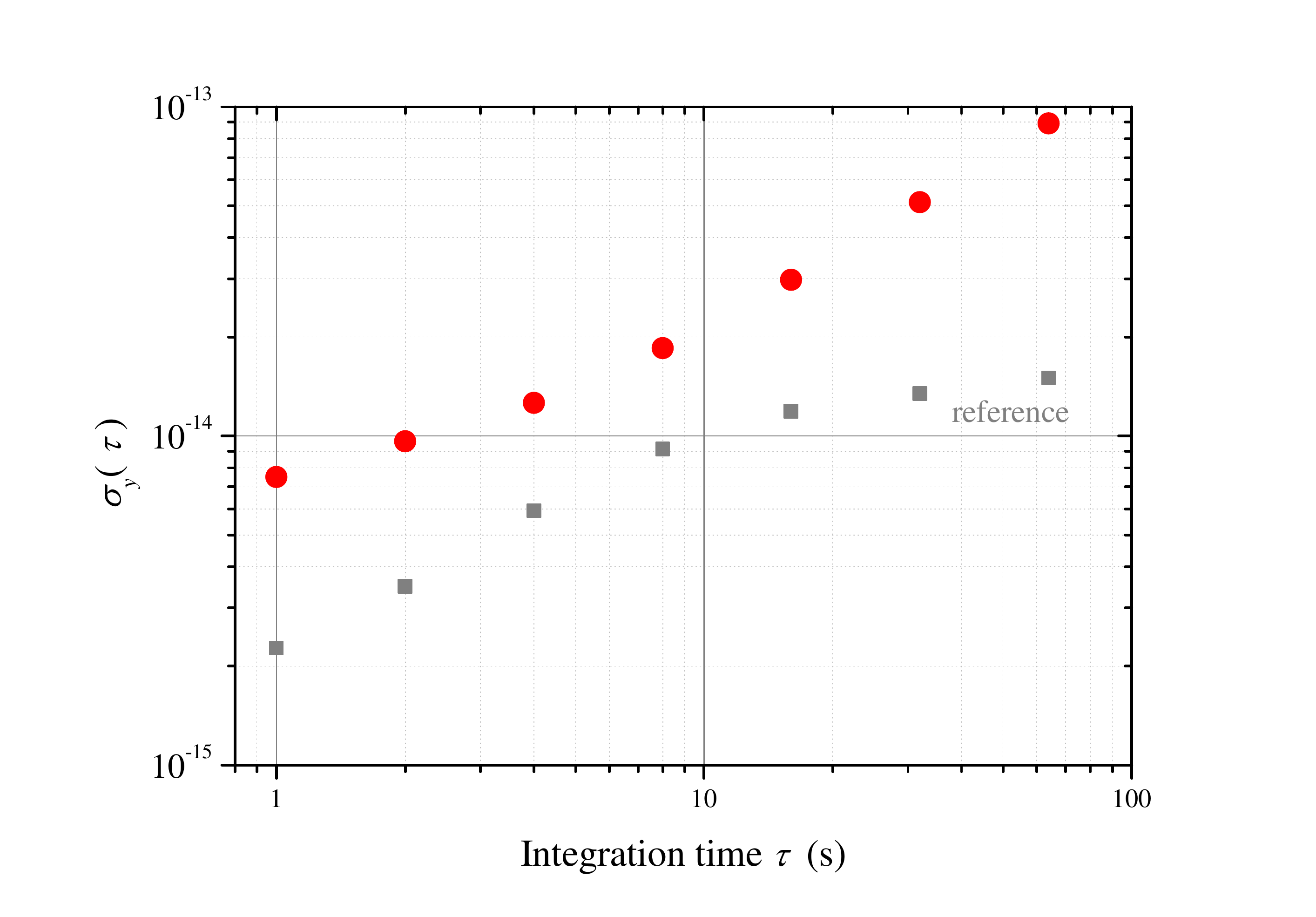}
	\caption{\textit{Allan deviation of the laser stabilized onto the ultra compact cavity (red circles). Grey squares represent the Allan Deviation of our reference laser stabilized on a spherical ULE cavity \cite{Didier2015}.}}
	\label{fig:stabcavite}
\end{figure}

The measured fractional frequency stability scales as $7.5 \times 10^{-15}\tau^{1/2}$ at short times, followed by a $2{\times}10^{-15}\tau$ behavior for $\tau>10$ s (Fig. \ref{fig:stabcavite}). This behavior is compatible with thermal fluctuations, and might stem from the high thermal sensitivity of our cavity, especially for the long-time drift. This will be fixed by the use of an INVAR cavity holder with lower temperature sensitivity. The fractional frequency instability measured at 1 s is consistent with the fractional frequency stability computed from the phase noise measured at $1$~Hz ($5 \times 10^{-15}$).

These performances are at the state-of-the art for such a short cavity; moreover, our simple design results in a setup with a reduced footprint and an overall volume of 30~L, interesting for field applications. 

\section{Conclusion}
In conclusion, we have developed and characterized a 25 mm-long pyramidal Fabry-Perot cavity with a ULE spacer, fused silica mirror substrates and crystalline mirror coatings. Thermal shields and regulation reject the influence of external temperature fluctuations just below $10^{-15}$. The acceleration sensitivity of the laser reaches $2{\times}10^{-10} \ \rm{m.s^{-2}}$ along the vertical axis. Even though this is compatible with the low $10^{-15}$ range when operating in a quiet laboratory environment, this sensitivity must be reduced in order to reach the cavity thermal noise limit in a noisier environment.

The cavity is equipped with highly- reflective crystalline coatings and exhibits the highest reported finesse (247 000) to this date using such coatings in an ultra-stable laser setup. Our ultra-stable laser fits in a volume of 30 liters (excluding electronics) and reaches a fractional frequency instability of $7.5{\times}10^{-15}$ at 1~s. Future upgrades will include a re-designed INVAR holder, for better rejection of both temperature and acceleration fluctuations, in order to reach the thermal noise floor around $1{\times}10^{-15}$.

\section{Acknowledgments}

This work was supported by the LABEX Cluster of Excellence FIRST-TF (ANR-10-LABX-48-01) and by the EQUIPEX OSCILLATOR-IMP (ANR 11-EQPX-0033), within the Program "Investissements d'Avenir” operated by the French National Research Agency (ANR), by the Council of the R\'egion de Franche-Comt\'e, and by the Centre National d'\'Etudes Spatiales (CNES).

The authors would like to thank G.D. Cole for the cavity finesse measurement.

\end{document}